\documentclass[prl,aps,twocolumn]{revtex4}
\usepackage{graphicx}
\usepackage{latexsym}
\usepackage{amsmath}
\usepackage{amssymb}
\begin{document}
\title{From inflation to late acceleration: A new cosmological paradigm}
\author{Supratik Pal \footnote{Electronic address: {\em{supratik\_v@isical.ac.in}}}}
${}^{}$
\affiliation{Physics and Applied Mathematics Unit,
Indian Statistical Institute,
203 B.T.Road, Kolkata 700 108, India}
\newcommand{\be}{\begin{equation}}
\newcommand{\ee}{\end{equation}}
\newcommand{\bea}{\begin{eqnarray}}
\newcommand{\eea}{\end{eqnarray}}
\newcommand{\bml}{\begin{subequations}}
\newcommand{\eml}{\end{subequations}}
\newcommand{\bfig}{\begin{figure}}
\newcommand{\efig}{\end{figure}}

\vspace{.5in}

\begin{abstract}

A new idea of deriving a cosmological term from an underlying theory has been proposed
in order to explain the expansion history of the universe.
We obtain the scale factor 
with this derived cosmological term and demonstrate that it reflects all the characteristics
of the expanding universe in different era so as to result in
 a transition from inflation to late acceleration through intermediate decelerating phases
by this single entity.
We further discuss certain observational aspects of this paradigm.

\end{abstract}


\maketitle


Explaining the expansion history of the universe is one of the leading problems facing
physicists over the years. Gravitational force, by its nature, is attractive.
But both Supernovae \cite{sn} and CMB data \cite{cmb} 
 suggest that the present universe is accelerating. 
Hence there must be some yet-unknown entity that supplies an effective `anti-gravity'.
Understanding the nature and evolution of this entity 
has drawn much attention from
the gravity as well as field theory communities. 
So far the most promising candidate is the cosmological constant $\Lambda$ \cite{cc}.
However, if cosmological constant is there, one needs extreme fine-tuning to match
the observationally tiny value of  $\rho_{\Lambda} \sim 10^{-124} GeV^{4}$
 for recent time from any underlying theory, which is still an open issue.
Subsequently, there are several alternative candidates.
Examples include
modifications of either the matter sector (quintessence, Kessence, Chaplygin gas etc. \cite{de})
or   the gravity sector
(Brans-Dicke theory \cite{bd}, $f(R)$ gravity \cite{fr}, DGP \cite{dgpde}
or generalized DGP-RS models \cite{brde}).
Each one of them has its own merits and demerits, 
which makes the issue further wide open.

On the other hand, inflationary scenario \cite{infl} suggests that the universe had
undergone a very fast accelerated expansion at early time, which  
requires a large cosmological constant. So, 
it appears nearly impossible to
unify inflation and late  acceleration under this common umbrella.
Precisely, there is no prima facie solution to the problem: 
why should an universal constant take vastly different values in two different epochs?
A possible wayout is that, 
maybe a variable cosmological term, whose value decays with time, can
serve the purpose, with the observational bound to its present value 
$\Lambda_0 \sim H_0^2$ \cite{vccold, vcc1, vcc2, vcc3, vcc4, vcc5}. 
There is a spectrum of possible models by taking the cosmological term
as a function of time, scale factor, Hubble parameter, deceleration parameter and so on
\cite{vccold, vcc1}. The observational constraints have also been investigated to some extent
\cite{vcc2, vcc3, vcc4}.
However, most of the models dealing with a variable cosmological term suffer from the drawback that
they are merely phenomenologically motivated, which is somewhat ad-hoc, lacking any strong theoretical argument,
thereby leading to serious debates like:   What is the underlying theory that
gives rise to its variable nature? Who governs the energy exchange between
a variable cosmological term and matter (or any other cosmic entity)?
Due to these serious drawbacks, the idea of a kinematical cosmological term 
did not appeal much of late and more popular models
with dynamical quantities like inflaton and dark energy came to the limelight.
The bottmeline is that, even with these dynamical models, the issue is not yet fully resolved.

In this article, we 
follow a  different route based on an underlying theory and demonstrate that this scenario indeed has great potentiality in addressing the  issue of acceleration of the universe in different era
from a common platform.
The reader may consider it as an altogether new paradigm rather than the continuation of 
an old idea.
We derive the form of the cosmological term from this theory,
which rules out any ad-hoc behavior.
This eventually answers to the question of energy exchange too.
The dramatic consequence of this idea is that it reflects all the characteristics
of the expansion history of the universe of all era by a single  entity
and demonstrates a  transition from inflationary phase to late accelerating phase
through radiation-dominated and matter-dominated era. 
We also show that $\Lambda$CDM falls as a subset of this framework, which leads to
further observational consequences.



The Friedmann equations with a variable cosmological term are given by \cite{vcc1}
\bea 
\left( \frac{\dot a}{a} \right)^2 =  \frac{8 \pi G}{3} \rho + \frac{\Lambda(t)}{3} - \frac{k}{a^2}
\label{frid1}\\
\frac{\ddot a}{a} = - \frac{4 \pi G}{3} \left(\rho + 3 p \right) + \frac{\Lambda(t)}{3}
\label{frid2}
\eea
supplemented by an effective conservation equation 
\be
\frac{d}{da} \left(\rho a^{3(1+w)}\right)  = - \frac{a^{3(1+w)}}{8 \pi G} \frac{d\Lambda}{da}
\label{cons} 
\ee
for a barotropic fluid with equation of state $w$.
Further, for a spatially flat universe ($k=0$),
the above equations can be recast into a single second order differential equation
(a special case of Riccati equation), 
by introducing a new variable $a(t) = \left[ x(t)\right]^{2/3(1+w)}$ as
\be
\frac{1}{\Lambda(t)} \frac{d^2x}{dt^2} - \alpha^2 x=0
\label{riccati} 
\ee
where $\alpha = \sqrt{3}(1+w)/2$.
The cosmological term we derive from our theory is found to  have the form 
\be
\Lambda(t)= \Lambda_{i} e^{-t/t_0}
\label{ct} 
\ee
We stress here that, unlike most of the earlier models,
this form of the cosmological term  is not an ad-hoc choice, rather, is an outcome of an
 underlying theory. We shall, however,
demonstrate this  later in this article.

Let us now engage ourselves in analyzing the 
evolution of the scale factor for different era.
First, for very early time $t \rightarrow 0$, the term (\ref{ct}) is practically a constant
$\Lambda \sim \Lambda_{i}$, which
is large enough so as to take care of inflationary expansion, so that Eq (\ref{riccati})
reduces to 
\be
\frac{d^2x}{dt^2} - \Lambda_{i} \alpha^2 x=0
\ee
which has the solution
\be
x(t) = C_0 e^{\sqrt{\Lambda_i}\alpha t}  \Rightarrow a(t) \propto
e^{\sqrt{\frac{\Lambda_i}{3}} t}
\ee
under the substitution $a(t) = \left[ x(t)\right]^{2/3}$ and $\alpha = \sqrt{3}/2$,
revealing the evolution of the universe during inflation.

For post-inflation era, the value of the cosmological term
is exponentially suppressed. 
In general, $ \forall ~t \nrightarrow 0$, Eq (\ref{riccati}), together
 with (\ref{ct}) takes the form
\be
e^{t/t_0}\frac{d^2x}{dt^2} - \Lambda_{i} \alpha^2 x=0
\ee
which can be solved  to give $x(t)$ as follows
\bea
x(t) &=& C_1 ~{\rm Bessel} I_0\left(2\sqrt{\Lambda_{i}} t_0 \alpha e^{-t/2t_0} \right) \nonumber \\
&+& C_2  ~{\rm Bessel} K_0\left(2\sqrt{\Lambda_{i}} t_0 \alpha e^{-t/2t_0} \right)
\label{soln} 
\eea
($C_1, C_2$ are arbitrary constants which can be determined from initial conditions)
from where one can readily obtain the scale factor by substituting back the variable
$a(t)$. 
Here $t_0$ is the time  at the onset of this behavior of the scale factor,
which is essentially small and introduces a new parameter in the theory.

One might wonder whether this apparently complicated expression encodes proper information
about the expanding universe. Below we answer to this question.
First note that
 Bessel$I_0$  and Bessel$K_0$ show opposite behavior while varying with the argument.
For a given form of the argument, one of them is a growing function whereas the other is a decaying one, 
a combination of which will naturally
give an initially decelerating and later accelerating phase.
\begin{figure}[t]
{\centerline{\includegraphics[width=6cm, height= 4cm] {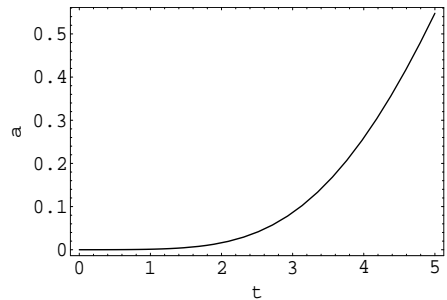}}}
\caption{Plot of the variation of the scale factor with time for a specific choice
of the parameters $C_1 =1, C_2 = 500, \sqrt{\Lambda_{i}} t_0 \alpha=5$. Here the horizontal axis
represents time and the vertical axis the scale factor.}
\end{figure}
FIG. 1 depicts the scale factor with the  entire function (\ref{soln}), which reveals  this behavior. Note that the parameters  satisfy the desired criteria that
this form of $K_0$ dominates with time.
The behavior of the scale factor will be further transparent if we analyze the limiting cases of
those functions. 
For small argument limit, the functions are given by: $I_0(y) \sim (y/2)^0/\Gamma(1) = 1$ 
and $K_0(y) \sim - \ln y$ so that we have 
\be
x(t) = C_1 -  C_2 \ln (2\sqrt{\Lambda_{i}} t_0 \alpha)
+ \frac{C_2}{2 t_0} t
\ee
The arbitrary constants $C_1$ and $C_2$ are chosen so that $C_1 =
C_2 \ln (2\sqrt{\Lambda_{i}} t_0 \alpha)$. This, of course, resonates with the values of 
the parameters  in FIG. 1, since $t_0$ suppresses the term in the parenthesis to a small value. 
By substituting $a(t) = \left[ x(t)\right]^{2/3(1+w)}$, we find that the
scale factor indeed behaves as:
\begin{itemize}
\item $a(t) \propto t^{1/2} \Rightarrow$ Radiation-dominated era 
\item $a(t) \propto t^{2/3} \Rightarrow$ Matter-dominated era  
\end{itemize}
which reveal exactly the cosmological evolution of the universe during decelerating phases.

What about the late accelerating phase? For large argument limit,
the functions behave as: 
$I_0(y) \sim (1/\sqrt{2 \pi y})e^{y}$ and $K_0(y) \sim (\sqrt{\pi/2y}) e^{-y}$
so that the scale factor is given by
\be
a(t) \propto \left[\frac{C_1}{\sqrt{2 \pi}} e^{y} + C_2 \sqrt{\frac{\pi}{2}}  e^{-y}\right]^{2/3}
y^{-1/3} 
\ee
with $y(t) = 2\sqrt{\Lambda_{i}} t_0 \alpha e^{-t/2t_0}$. This is 
clearly a generalization of the usual 
$a(t) \propto [e^{\frac{3}{2}\sqrt{\frac{\Lambda_0}{3}}t} + e^{-\frac{3}{2}\sqrt{\frac{\Lambda_0}{3}}t}]^{2/3}
=[\sinh(\frac{3}{2}\sqrt{\frac{\Lambda_0}{3}}t)]^{2/3}$
obtained from $\Lambda$CDM model, and thus, 
includes $\Lambda$CDM as a subset.
Consequently, all the features of $\Lambda$CDM are preserved within this framework.
 
We thus arrive at the following remarkable conclusion:
{\em All the informations invloving the scale factor in different era
are encoded in this framework
so that the expansion history of the universe is reflected by a single dynamical entity
derived from an underlying theory}.

The situation is further interesting from observational ground. 
One can calculate quantities  such as the Hubble parameter $H$,
deceleration parameter $q$, age of the universe, luminosity distance \cite{de},
 statefinder parameters $\{r, s\}$ \cite{statefinder} and the very recently introduced
$Om(z)$ and acceleration probe $\bar q$ \cite{newprobe}, and confront them with observations.
 For example, the Hubble parameter is given by
\be
H (t)= \frac{\dot a}{a} = \frac{2}{3} \frac{C_1 I_0^{'} +C_2 K_0^{'}}{C_1 I_0 +C_2 K_0} \frac{dy}{dt}
\label{hub1} 
\ee
where a prime denotes a derivative w.r.t. $y$.
In large argument limit, $I_0^{'} \sim (1/\sqrt{2 \pi y})e^{y} \sim I_0$ 
and $K_0^{'} \sim - (\sqrt{\pi/2y}) e^{-y} \sim -K_0$ and this form of $K_0$ dominates at late time,
so that Eq (\ref{hub1}) is simplified to
\be
H (t) \approx \sqrt{\frac{\Lambda_i}{3}} e^{-t/2t_0} 
\label{hub2}
\ee
The exponential factor suppresses the cosmological term to a tiny value 
(clearly not exactly zero)
during late time
so that the observed bound to its present value $\Lambda_0 \sim H_0^2$ is satisfied.
Likewise, the other quantities can be derived and confronted with observations.



We shall now concentrate on a possible answer to the question about the origin
of such a cosmological term. 
Below we  demonstrate that there is indeed an underlying theory, namely,
 braneworld gravity, that gives rise to an entity behaving as a cosmological term 
of the form given in Eq (\ref{ct}) and, at the same time, takes care of the energy exchange
issue.

In the general  
braneworld scenario, our 4D universe is a subspace (brane) of a higher dimensional
Vaidya-anti de Sitter space (bulk), which 
exchanges energy with the brane
\cite{maartrev, maartbulk, langrev, lang1, chamb,  vads5gen, maartprog, ansatz,  sustr, sucol, suden}.
To a brane-based (4D) observer, this energy exchange results in a non-conservation 
equation for brane matter 
\cite{maartrev, langrev,  sustr, sucol, suden}.
\be
\dot\rho + 3 \frac{\dot a}{a} (\rho + p) = -2 \psi
\label{noncons} 
\ee
In this scenario, the (4D) Friedmann equations on the brane are given by
\cite{sucol, suden}
\bea
\left( \frac{\dot a}{a} \right)^2 =  \frac{8 \pi G}{3}   \left(\rho +
\frac{\rho^2}{2\lambda} \right) + {\cal E} - \frac{k}{a^2}
\label{fridbr1} \\
\frac{\ddot a}{a} = -\frac{4 \pi G}{3} \left[(\rho +3p) + \frac{(\rho^2+ 6\rho p)}{2\lambda} \right] - {\cal E} - \frac{\kappa_5^2}{3}\psi
\label{fridbr2}
\eea
 where ${\cal E}$ and $\psi$ are contributions from  bulk geometry (Weyl tensor) and bulk radiation field on the brane.
Note that we have written the above equations in the RS gauge, i.e., without introducing any  arbitrary cosmological constant. In what follows we shall show that it rather appears 
from  the underlying theory.

The quadratic terms here are from brane matter/ radiation sector and hence, 
can be dropped from the above equations
since brane tension $\lambda$ dominates over matter/ radiation right from the era they are formed.
The quadratic terms are relevant only if one 
considers $\rho$ to be inflaton field. We are not considering any inflaton either, 
and thus they are
irrelevant during inflation too. Consequently, 
Eqs (\ref{fridbr1}) and (\ref{fridbr2}) go over to Eqs
(\ref{frid1}) and (\ref{frid2}) respectively under the following identification of terms
\be
{\cal E} =  - {\cal E} - \frac{\kappa_5^2}{3}\psi
= \frac{\Lambda(t)}{3}
\label{cond} 
\ee
This leads to the most crucial feature of the scenario.
It is well-known that for any decaying cosmological term model, formation of matter  at the cost of 
the cosmological term is governed by Eq (\ref{cons}) giving \cite{vccold, vcc2} 
\be
\dot\rho + 3 \frac{\dot a}{a} (\rho + p) = - \dot \Lambda
\label{decay} 
\ee
Eq (\ref{decay}), together with Eq (\ref{noncons}) and the expression for the radiation field
$\psi$  obtained from Eq (\ref{cond}),  gives
\be
\dot \Lambda = - \left(4/\kappa_5^2\right)  \Lambda
\ee
which has the solution of the form
\be
\Lambda (t) = A e^{-(4/\kappa_5^2)t}
\ee
immediately leading to the exact form of $\Lambda$ used in Eq (\ref{ct}),
with  $A = \Lambda_i$, $t_0 = (\kappa_5^2/4)$.
This gives a possible explanation of why we started with such a behavior for
$\Lambda$, and subsequently, provides an underlying theory behind its origin.
 Nevertheless, as obvious from the discussions preceding Eq (\ref{noncons}),
 the energy exchange is inbuilt in the theory, so that it gives a natural explanation
of the decay of the cosmological term to the other cosmic entities via braneworlds.


Let us now summarize the key features of this article. The present
article has the following significant contributions: 
\begin{enumerate}
\item The exact solution for the scale factor has been  obtained by using a cosmological term.

\item The  form of the cosmological term has been derived from an underlying theory, which rules out
ad hoc behavior and explains energy exchange too.  

\item Transition from inflation to late acceleration, via radiation-dominated
and matter-dominated era, has been demonstrated.

\item $\Lambda$CDM is found to be a subset of this framework and thus,  all of its 
features are preserved.

\item Certain observational aspects have also been addressed.

\end{enumerate}

This framework opens up a spectrum of possible avenues to venture. 
For example,
the effect of this decaying cosmological term  on reheating can be investigated.
Cosmological perturbations and behavior of power spectrum can also be studied.
Another issue is to explore the observational sector,  
discussed to some extent in the present article, which may further comment on
its acceptability as the driving entity for the expansion of the universe.
We end up with this optimistic note 
that a details analysis of the observational sector may reveal its credentials in future.



\begin{references}

\bibitem{sn} S. Perlmutter et al., Astrophys. J. {\bf 517} (1999) 565

\bibitem{cmb} D. N. Spergel et al., Astrophys. J. {\bf 148} (2003) 175 

\bibitem{cc} S. M. Carroll, Living Rev. Relativity {\bf 3} (2001) 1;
T. Padmanabhan, Phys. Rept. {\bf 380}, 235 (2003)

\bibitem{de} V. Sahni, Lect. Notes Phys. \textbf{653} (2004) 141;
E. J. Copeland, M. Sami and S. Tsujikawa, Int. J. Mod. Phys. {\bf D 15}, 1753 (2006);
J. Frieman, M. Turner  and D. Huterer, arXiv:0803.0982 [astro-ph]

\bibitem{bd} S. Sen and A. A. Sen, Phys. Rev. {\bf D 63} (2001) 124006;
A. A. Sen, S. Sen and S. Sethi, Phys. Rev. {\bf D 63} (2001) 107501

\bibitem{fr} S. Das, N. Banerjee and N. Dadhich, Class. Quant. Grav. {\bf 23}, 4159 (2006);
N. J. Poplawski, Class. Quant. Grav. {\bf 23}, 2011 (2006);
L. Amendola, R. Gannouji, D. Polarski and S. Tsujikawa, Phys. Rev. {\bf D 75}, 083504 (2007)

\bibitem{dgpde} R. A. Brown, R. Maartens, E. Papantonopoulos and V. Zamarias, JCAP {\bf 11} (2005) 008

\bibitem{brde} V. Sahni and Y. Shtanov, JCAP {\bf 11} (2003) 014;
V. Sahni, Y. Shtanov and A. Viznyuk,  JCAP {\bf 12}, (2005) 005;
Y. Shtanov, A. Viznyuk and V. Sahni, Class. Quant. Grav. {\bf 24} (2007) 6159;

\bibitem{infl} A. R. Liddle and D. H. Lyth, {\em Cosmological Inflation and Large-scale Structure},
Cambridge University Press, Cambridge (2000)

\bibitem{vccold} M. Gasperini, Phys. Lett. {\bf 194 B}, 347 (1987);
J. M. Overduin, P. S. Wesson and S. Bowyer, Astrophys. J. {\bf 404}, 1 (1993);
J. C. Carvalho, J. A. S. Lima and I. Waga, Phys. Rev. {\bf D 46}, 2404 (1992);
D. Pavon, Phys. Rev. {\bf D 43}, 375 (1991);
M. S. Berman, Phys. Rev. {\bf D 43}, 1075  (1991);
A. Beesham, Phys. Rev. {\bf D 48}, 3539 (1993);
J. A. S. Lima and M. Trodden, Phys. Rev. {\bf D 53} 4280 (1996);
V. Sahni and A. Starobinsky, Int. J. Mod. Phys. {\bf D 9} (2000) 373


\bibitem{vcc1} J. M. Overduin and F. I. Cooperstock, Phys. Rev. {\bf D 58}, 043506 (1998)

\bibitem{vcc2} S. Carneiro, C. Pigozzo, H. A. Borges and J. S. Alcaniz, 
Phys. Rev. {\bf D 74} (2006) 023532

\bibitem{vcc3} S. Carneiro, M. A. Dantas, C. Pigozzo and J. S. Alcaniz,
Phys. Rev. {\bf D 77}, 083504 (2008)

\bibitem{vcc4} Y. Z. Ma, arXiv:0708.3606 [astro-ph]

\bibitem{vcc5} H. A. Borges, S. Carneiro, J. C. Fabris and C. Pigozzo;
Phys. Rev. {\bf D 77}, 043513 (2008);
Y. Gong and X. Chen, Phys. Rev. {\bf D 77}, 103511 (2008)

\bibitem{statefinder} V. Sahni, T. D. Saini, A. A. Starobinsky and U. Alam,
JETP Lett. {\bf 77} (2003) 201; Pisma Zh. Eksp. Teor. Fiz. {\bf 77} (2003) 249;
U. Alam, V. Sahni, T. D. Saini and A. A. Starobinsky,
Mon. Not. Roy. Astron. Soc. {\bf 344} (2003) 1057

\bibitem{newprobe} V. Sahni, A. Shafieloo and A. A. Starobinsky,
arXiv:0807.3548 [astro-ph]



\bibitem{maartrev} R. Maartens, Living Rev. Relativity {\bf 7} (2004) 7 

\bibitem{langrev} D. Langlois, Prog. Theor. Phys. Suppl. {\bf 148} (2003) 181


\bibitem{maartbulk} E. Leeper, R. Maartens and C. Sopuerta, Class. Quant.
Grav. {\bf 21} (2004) 1125

\bibitem{lang1} D. Langlois, L. Sorbo and M. Rodriguez-Martinez, Phys. Rev. Lett. {\bf 89} (2002) 171301;
D. Langlois and L. Sorbo, Phys. Rev. {\bf D68} (2003) 084006; 

\bibitem{chamb}  A. Chamblin, A. Karch and A. Nayeri, Phys. Lett. {\bf B509} (2001) 163;
D. Langlois and L. Sorbo, Phys. Rev. {\bf D68} (2003) 084006

\bibitem{vads5gen} L. A. Gergely, Phys. Rev. {\bf D68} (2003) 124011;
  L. A. Gergely, E. Leeper and R. Maartens, Phys. Rev. {\bf D70} (2004) 104025;
I. R. Vernon and D. Jennings, JCAP {\bf 07} (2005) 011;
 D. Jennings, I. R. Vernon, A. C. Davis and C. van de Bruck, JCAP {\bf 04} (2005) 013;
L. A. Gergely and Z. Keresztes, JCAP {\bf 01} (2006) 022;

\bibitem{maartprog} R. Maartens, Prog. Theor. Phys. Suppl. {\bf 148}, 213 (2003)


\bibitem{ansatz} K. Maeda, Lect. Notes Phys. {\bf 646} (2004) 323;
 D. Langlois and M. Rodriguez-Martinez, Phys. Rev. {\bf D64} (2001) 123507;
 G. Kofinas, G. Panotopoulos and T. N. Tomaras, JHEP {\bf 01} (2006) 107;


\bibitem{sustr} S. Pal, Phys. Rev. {\bf D 74}, 024005 (2006) 

\bibitem{sucol} S. Pal, Phys. Rev. {\bf D 74}, 124019 (2006) 

\bibitem{suden} S. Pal, Phys. Rev. {\bf D 78}, 043517 (2008)  [arXiv: 0806.2505 [gr-qc]]





\end{references}
\end{document}